\documentclass[dvips,twoside,fleqn]{article}
\usepackage{times}
\usepackage{epsfig}
\usepackage{amstext}
\usepackage{amssymb}
\usepackage{espcrc2}

\title{$(\lambda \Phi^4)_4$ theory on the lattice: evidence for a non-trivial rescaling 
of the scalar condensate.}
\author{
P. Cea\address{INFN - Sezione di Bari - Via Amendola 173 - 70126 Bari - Italy},
M. Consoli\address{INFN - Sezione di Catania - Corso Italia 57 - 95129 Catania - Italy},
and L. Cosmai$^{\text{a}}$
}

\begin{document}

\begin{abstract}
A lattice simulation in the broken phase of $(\lambda \Phi^4)_4$  theory in the Ising limit 
suggests that, in the
continuum limit, the scalar condensate rescales by a factor different from the conventional
wavefunction renormalization. Possible effects on the present bounds of the Higgs mass are
discussed.
\end{abstract}
\maketitle

\section{INTRODUCTION}

It is widely believed that \cite{aizen,froh,sokal,latt,luscher87,glimm,book}
$(\lambda\Phi^4)_4$ theories are ``trivial''.
The conventional interpretation is 
based on leading-order
Renormalization-Group-Improved-Perturbation-Theory (RGIPT). 
However a quite different interpretation is advocated in Refs. 
\cite{zeit}.  
A key feature of the alternative picture is the presence 
of a non-trivial rescaling of the ``renormalized'' vacuum field:
\begin{equation}
\label{vR}
v_R\equiv v_B/\sqrt{Z_{\varphi}} \,.
\end{equation}
The role of $Z_{\varphi}$ is essential. It provides the key ingredient 
to get a non trivial effective potential in a ``trivial'' theory\cite{zeit,agodi1,agodi2}.
It turns out that\cite{rit2,V2}  in the continuum
limit ($\Lambda \to \infty$) 
\begin{equation}
\label{Zphi}
           Z_{\varphi} \sim \ln {{\Lambda}\over{M_h}} \to \infty  \,,
\end{equation}
so that, although $M^2_h/v^2_B \to 0$, one finds 
\begin{equation}
\label{vR2}
              {{M^2_h}\over{v^2_R}}=~ \Lambda-{\rm independent} \,.
\end{equation}
On the other hand in the continuum limit $Z_{\rm prop} \to 1$ consistently
with the trivial nature of the shifted field.
In order to directly test the prediction that 
$Z_{\varphi}$ differs from $Z_{\rm prop}$,  we present the results of a 
lattice simulation of the theory (in the Ising limit) where we compute 
the mass  and the residue $Z_{\rm prop}$ from a 2-parameter fit to 
the lattice data for the shifted-field propagator.  We then compute the 
zero-momentum susceptibility 
\begin{equation}
\label{phiB}
\left. 
{{1}\over{\chi}}=
\frac{d^2V_{\rm eff}}{d\varphi^2_B} \right|_{\varphi_B=\pm v_B}
\end{equation}
and hence obtain the dimensionless quantity
\begin{equation}
\label{m2chi}
Z_{\varphi} \equiv M^2_h \chi.  
\end{equation}
Finally, we compare $Z_{\varphi}$ with $Z_{\rm prop}$.

\section{NUMERICAL SIMULATIONS}

The one-component $(\lambda\Phi^4)_4$ 
theory   
\begin{eqnarray}
\label{action}
\lefteqn{
S =\sum_x \left\{ \frac{1}{2} \sum_\mu \left[ \Phi(x+\hat e_{\mu}) - 
\Phi(x) \right]^2 + \right. 
}  \nonumber  \\ 
\lefteqn{
\left. \qquad  \qquad \frac{r_0}{2} \Phi^2(x)  +
\frac{\lambda_0}{4} \Phi^4(x) -  J \Phi(x)  \right\}
}
\end{eqnarray}
becomes in the Ising limit 
\begin{eqnarray}
\label{ising}
\lefteqn{
S_{\rm Ising} = -\kappa
\sum_x\sum_{\mu} \left[ 
\phi(x+\hat e_{\mu})\phi(x) + \right.}
\nonumber \\
\lefteqn{
\left. \qquad  \qquad \qquad  \qquad  \qquad
\phi(x-\hat e_{\mu})\phi(x) \right]
}    
\end{eqnarray}
with $\Phi(x)=\sqrt{2\kappa}\phi(x)$ and $|\phi(x)| = 1$. 

The shifted field propagator, defined at 
$p_{\mu} \neq 0$, can be computed 
as 
\begin{equation}
G(p)=\langle \sum_x \exp (ip x) h(x)h(0)\rangle
\end{equation}
for the values
$p_{\mu}={{2\pi}\over{L}}n_{\mu}$ with $n_{\mu}\neq 0 $.
An excellent fit to the lattice data is obtained 
by using the 2-parameter formula
\begin{equation}
G(p)= {{Z_{\rm prop}}\over{\hat{p}^2 + m^2_{\rm latt}} }
\end{equation}
where $m_{\rm latt}$ is the dimensionless lattice mass and 
$\hat{p}_{\mu}=2 \sin {{p_{\mu} }\over{2}}$  (see Fig.1).
%
%
%
\begin{figure}[t]
\label{Fig1}
\begin{center}
\epsfig{file=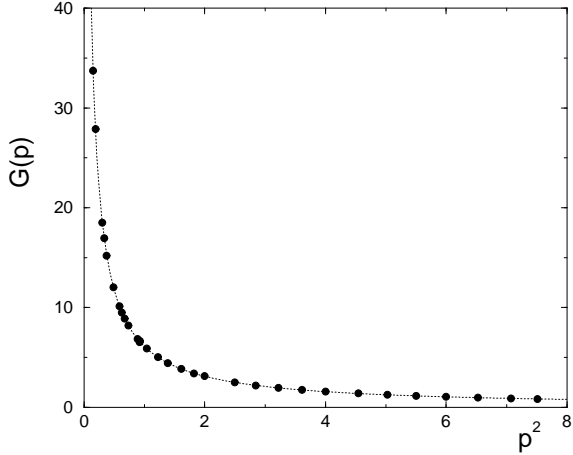,width=7.5truecm}
\caption{ The lattice data for the propagator (Eq.(8)) (circles) 
at $\kappa=0.07510$ on a $32^4$ lattice with
superimposed the fit Eq.(9) (dotted line).}
\end{center}
\end{figure}

The susceptibility $\chi$ is measured directly as
\begin{equation}
\label{suscep}
\chi_{\rm latt}=L^4 \left[ \left\langle \Phi^2 \right\rangle - 
\left\langle \Phi \right\rangle^2 \right] \
\end{equation}
with $\Phi$ the average field for each lattice configuration.
Moreover we define
\begin{equation}
\label{Zlatt}
Z_{\varphi} \equiv m^2_{\rm latt}~ \chi_{\rm latt}  \,.
\end{equation}
To update our field configurations we used the Swendsen-Wang 
\cite{SW} cluster algorithm on $20^4$, $24^4$ and $32^4$ lattices.
 After discarding 10K sweeps for thermalization,
we have performed 50K sweeps, measuring our observables every 5 sweeps.
We have computed at different values of the hopping parameter $\kappa$ in 
order to obtain 
a correlation length $\xi_{\rm latt}=1/m_{\rm latt}$ 
in the range 2 to $L/4$. The upper limit of
the correlation length is required in order to avoid finite-size 
effects \cite{montvay,jansen}. 

Our results
for $Z_{\varphi}$ and $Z_{\rm prop}$, in the broken phase 
 are reported in Fig.2, and show a sizeable
difference for $m_{\rm latt} <0.3$. 
%
%
%
\begin{figure}[t]
\label{Fig2}
\begin{center}
\epsfig{file=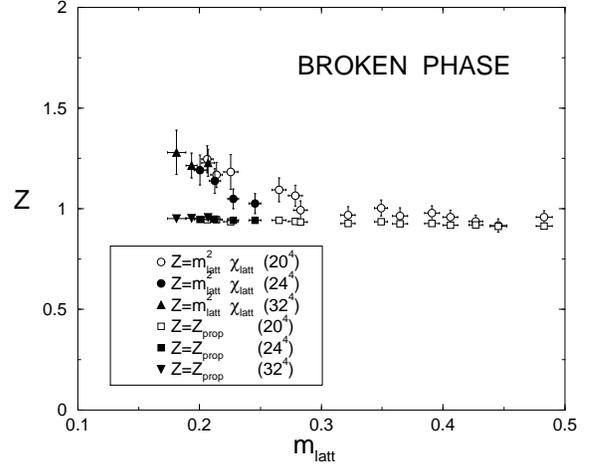,width=7.5truecm}
\caption{$Z_\varphi$  and $Z_{\text prop}$  in the broken phase 
versus $m_{\text latt}$.}
\end{center}
\end{figure}

We have performed a consistency 
check that no such effect is 
present in the symmetric phase (Fig.3). 
%
%
%
\begin{figure}[t]
\label{Fig3}
\begin{center}
\epsfig{file=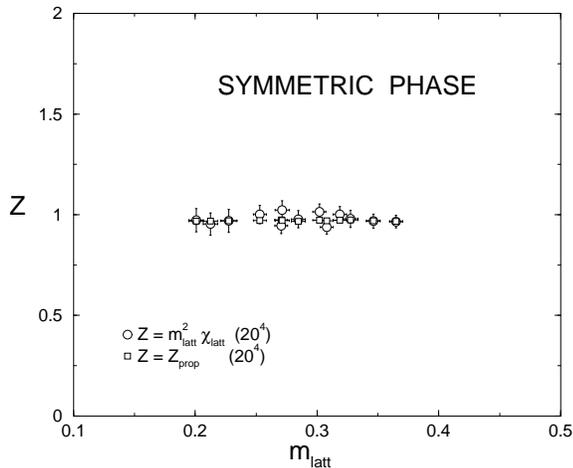,width=7.5truecm}
\caption{$Z_\varphi$  and $Z_{\text prop}$  in the symmetric phase 
versus $m_{\text latt}$.}
\end{center}
\end{figure}

As an additional check, we have compared 
with available data in the literature\cite{montvay,jansen} 
 both in the symmetric and broken phase 
and found good agreement. 

\section{CONCLUSIONS}

Our numerical 
simulation of $(\lambda\Phi^4)_4$, in the Ising limit, 
shows a clear difference between two {\it measured} 
quantities: the rescaling of
the ``condensate'' $Z_{\varphi}$ 
and the more conventional quantity
$Z_{\rm prop}$ associated with the residue of the shifted field propagator.
The effect shows up when increasing the correlation length and should become 
more and more important by approaching the continuum limit
of quantum field theory $m_{\rm latt} \to 0$.
Therefore, the relation of the lattice vacuum field $\langle\Phi\rangle$ to
the Fermi constant
and the same limits on the Higgs mass can sizeably be affected. Indeed, these
have been based on the quantity \cite{lang}
\begin{equation}
\label{rprop}
R_{\rm prop}= {{ m_{\rm latt}} \over{ \langle\Phi\rangle }}~
\sqrt{Z_{\rm prop}} 
\end{equation}
rather than 
\begin{equation}
\label{rphi}
R_{\varphi}= {{ m_{\rm latt}} \over{ \langle\Phi\rangle } }~
\sqrt{Z_{\varphi}}  \,.
\end{equation}
The discovery of $Z_{\varphi}$ requires
a ``second generation'' of lattice simulations to re-check the
scaling behaviour of the various quantities 
and compare with all available theoretical descriptions of the continuum limit.

\end{document}